\documentclass[aps,twocolumn]{revtex4}
\usepackage{graphicx,epsfig}
\begin{document}
\draft
\preprint{}
\title{Nuclear Structure corrections for Parity Non-conservation
in Atomic Ba$^+$ and Ra$^+$}
\author{P. K. Panda and B.P. Das}
\affiliation{Indian Institute of Astrophysics, Koramangala,
Bangalore-560034, India}
\begin{abstract}
We calculate the binding energy, charge radii in
$^{126}$Ba--$^{140}$Ba and $^{214}$Ra--$^{228}$Ra using the
relativistic mean field theory which includes scalar and vector
mesons. We then evaluate the nuclear structure corrections to the
weak charges for a series of isotopes of these atoms using different
parameters and estimate their uncertainty in the framework of this
model. Our results will have important implication for the ongoing
and planned parity non-conservation experiments and atomic structure
calculations on Ba$^+$ and Ra$^+$.
\end{abstract}
\maketitle

Parity non-conservation (PNC) in heavy atoms have provided an
important confirmation \cite{bouch1,edwards,wood} of the
SU(2)$\times$U(1) electro-weak sector of the Standard Model (SM). By
combining the results of precision measurements and calculations
using sophisticated many-body methods \cite{dzuba,blundell,bennett},
it is possible to extract the nuclear weak charge and compare with
its corresponding value in the SM. A discrepancy between these two
values could reveal the possible existence of new physics beyond the
SM \cite{lang}.

As first pointed out by Bouchiat and Bouchiat \cite{bouch1}, the
matrix element of the PNC Hamiltonian scales as $Z^3$. It is
primarily because of this reason that heavy atoms are considered to
be the best candidates for PNC experiments. A high precision
measurement of PNC in atomic cesium \cite{wood} has reduced
significantly the uncertainty ($<$ 1\%) in the determination of the
nuclear weak charge, $Q_W$, of the Cs nucleus and the deviation from
the SM is about $1\sigma$ \cite{dzuba02}. It would also be desirable
to consider other systems that have the potential to yield accurate
values of the nuclear weak charge. Ba$^+$ and Ra$^+$ deserve special
mention in this context. The transition of interest are
$6s^2S_{1/2}\rightarrow 5d^2D_{3/2}$ for Ba$^+$ and
$7s^2S_{1/2}\rightarrow 6d^2D_{3/2}$ for Ra$^+$. An experiment is
underway for Ba$^+$ using the techniques of ion trapping and laser
cooling and another has been proposed for Ra$^+$
\cite{fortson93,koerber03}. Relativistic many-body calculations
\cite{sahoo,dzuba01} have also been carried out on these two ions.

The experimental result needs input from atomic structure
calculations involving the interplay of electromagnetic and weak
interactions. However, the small but non-negligible effects of
nuclear size must be addressed before an interpretation of PNC data
in terms of the fundamental electro-weak couplings is possible. Thus
nuclear structure could become a crucial factor in the
interpretation of PNC experiments of increasing accuracy
\cite{wieman,fortson,wilets,chen,panda}. An extensive discussion on
the sensitivity of atomic PNC and electric dipole moments to
possible new physics has been recently reported in Ref.
\cite{ginges}.

There have been earlier studies to determine nuclear structure
effects in PNC in atomic cesium using non-relativistic potentials
\cite{wilets,chen} as well as relativistic models \cite{panda}. In
this letter we present a relativistic calculation of these effects
for the Ba and Ra isotopes using the relativistic mean field theory
(RMF). It is motivated by the current efforts to observe PNC in
Ba$^+$ and Ra$^+$.

The RMF theory first proposed by Teller and co-workers
\cite{Teller,Johnson,Duerr} and later by Walecka \cite{Walecka} and
developed by others has been fairly successfully applied to both
nuclear matter and finite nuclei. The method gives good description
for binding energies, root mean square (rms) radii, quadruple and
hexadecapole deformations and other nuclear properties not only for
the spherical, but also for the deformed nuclei. The same parameter
set of the model also describes well the properties of nuclear
matter. One of the major attractive features of the RMF approach is
the incorporation of the spin-orbit interaction due to to the
presence of the one body Dirac Hamiltonian and the nuclear shell
structure automatically arises from the nucleon-nucleon interaction
via the scalor and vector mesons. We can therefore expect the RMF
calculation to provide useful information on nuclear structure
corrections to atomic PNC.

In the Standard Model, the electron-nucleon interaction is mediated
by both the photon and the intermediate boson $Z^0$. The energy
involved in the atomic PNC experiments are usually only a fraction
of an eV, while the mass of the $Z^0$ is $\simeq 92$ GeV, and so the
parity non-conserving interaction may be written as a contact
interaction. We have
\begin{eqnarray}
H_{pnc}&=&\frac{G_F}{\sqrt{2}}\sum_{eB}\big[
C_{1B}\int\psi_B^\dagger \psi_B\psi_e^\dagger\gamma_5\psi_e d^3
r\nonumber\\
&+&C_{2B}\int\psi_B^\dagger {\bf \sigma}_B\psi_B\cdot
\psi_e^\dagger {\bf a} \psi_e d^3 r\big],
\end{eqnarray}
where $B$ stands for $n$ (neutron) or $p$ (proton). The first term grows
coherently with nucleon numbers $N$ and $Z$. The second term together
with the anapole moment term amounts to at most a few percent of the
first term in heavy atoms. We shall therefore consider only the first
term. The effective Hamiltonian becomes
\begin{equation}
H_{pnc}=\frac{G_F}{\sqrt{2}}\int\big[NC_{1n}\rho_n({\bf r})+ZC_{1n}
\rho_p({\bf r})\big] \psi_e^\dagger\gamma_5\psi_e d^3 r,
\end{equation}
where the proton and
neutron densities, $\rho_{p,n}({\bf r})$, are
normalized to unity. We have assumed the Standard Model nucleon couplings
\begin{equation}
C_{1p}\equiv2C_{1u}+C_{1d}=\frac{1}{2}(1-4\sin^2\theta_W),
\end{equation}
\begin{equation}
C_{1n}\equiv2C_{1u}+C_{1d}=-\frac{1}{2}.
\end{equation}
We need the spatial variation of the electron part
$\psi_e^\dagger\gamma_5\psi_e$ over the nucleus, its normalization
and its dependence on nuclear structure. PNC effects are dominated by
$s$-electrons $(\kappa=-1)$ coupled to $p$-electrons $(\kappa=1)$.
This can be expressed as
\begin{equation}
\rho_5(r)\equiv\psi_p^\dagger\gamma_5\psi_s=C(Z){\cal N}(Z,R)f(r),
\end{equation}
where $C(Z)$ contains all atomic-structure effects for a point nucleus
including many-body correlations,
${\cal N}\equiv\psi_p^\dagger(0)\gamma_5\psi_s(0)$ is the normalization
factor for single electron and $f(r)$ describes the spatial variation
[normalized such that $f(0)=1$]. It is the integrals
\begin{equation}
q_{n,p}=\int f(r)\rho_{n,p}(r)d^3 r,
\end{equation}
which determine the effect of the proton and neutron distributions on
the PNC observables.

PNC effects are proportional to the matrix element between two
atomic states $i$ and $j$,
\begin{equation}
{\cal O}=\langle i|H_{pnc}|j\rangle=\frac{G_F}{2\sqrt{2}}C(Z){\cal
N} \big[Q_W(N,Z)+Q_W^{nuc}(N,Z)\big],
\end{equation}
where $Q_W(N,Z)$ is the weak charge. For the Standard Model, the weak
charge takes the form at tree level as
\begin{equation}
Q_W(N,Z)=-N+Z(1-4\sin^2\theta_W).
\label{qw}
\end{equation}
The nuclear structure correction $Q_W^{nuc}(N,Z)$ describes the part
of the PNC effect that arises from the finite nuclear size. In the
same approximation as (\ref{qw}) above
\begin{equation}
Q_W^{nuc}(N,Z)=-N(q_n-1)+Z(1-4\sin^2\theta_W)(q_p-1).
\label{qwn}
\end{equation}

The proton (charge) nuclear form factors needed for $q_p$ and ${\cal N}$
are generally well known from the measurements of the charge distribution of
nuclei close to the stable valley and many unstable nuclei as well.
The neutron nuclear form factor needed for $q_n$ is not well determined
experimentally and is model dependent.
To estimate the importance of PNC in nuclear structure, the form factor
can be approximated to the order of $(Z\alpha)^2$ for a sharp nuclear
surface, and neglecting the electron mass in comparison
with the nuclear Coulomb potential \cite{fortson},
\begin{equation}
f(r)\simeq 1- \frac{1}{2}(Z\alpha)^2[(r/R)^2-\frac{1}{5}(r/R)^4+
\frac{1}{75}(r/R)^6].
\label{form}
\end{equation}
In the above, for a sharp nuclear surface density distribution, the only
relevant parameter is the nuclear radius $R$ and $\langle r^{2n}\rangle =
3/(2n+3)R^{2n}$.

One of the motivations for further improving atomic PNC experiments
is to test the Standard Model parameters. After the inclusion of
radiative corrections, we begin by rewriting Equation (8) and (9) in
the form
\begin{eqnarray}
Q_W(N,Z)&=&0.9878\times[-N+Z(1-4.0118\bar
x)]\nonumber\\
&&\hspace{0.8in}\times(1.0+0.00782T),
\end{eqnarray}
\begin{equation}
{\bar x}=0.23124\pm 0.00017+0.003636S-0.00258T,
\end{equation}
where $\bar x$ is assumed here to be defined at the mass scale $M_Z$
by modified minimal subtraction \cite{lang}, $S$ is the
parameter characterizing the isospin conserving new quantum loop
corrections and $T$ characterizing isospin breaking corrections. The
nuclear structure correction to $Q_W$ is given by
\begin{equation}
Q_W^{nuc}(N,Z)=0.9878\times[-N(q_n-1)+Z(1-4.0118\bar x)(q_p-1)]
\end{equation}
The coefficients $q_{n,p}$ defined earlier in equation (6) contain
the nuclear structure effects. We have included the intrinsic
nucleon structure contributions in evaluating the nuclear structure
correction. We use \cite{wilets}
\begin{equation}
q_{p,n}=\int d^3 r \left [ \rho_{p,n} (r) + \frac{1}{6} \langle r^2
\rangle_{I,(p,n)} \Delta^2 \rho_{p,n}/Q^W_{p,n}\right ] f(r)
\end{equation}
where  $\langle r^2 \rangle_{I,(p,n)}$ are the nucleon weak radii
and $Q^W_{p,n}$ are nucleon weak charges.

The relativistic Lagrangian density for a nucleon-meson many-body
system \cite{Set,Gam}
\begin{eqnarray}
{\cal L}&=&\bar\psi_i(i\gamma^\mu\partial_\mu-M)\psi_i\nonumber\\
&+&{1\over 2}\partial^\mu\sigma\partial_\mu\sigma-
{1\over 2}m_\sigma^2\sigma^2+{1\over 3}g_2\sigma^3
+{1\over 4}g_3\sigma^4 -g_s\bar\psi_i\psi_i\sigma\nonumber\\
&-&{1\over 4}\Omega^{\mu\nu}\Omega_{\mu\nu}
+{1\over 2}m_\omega^2 \omega^\mu \omega_\mu
+{1\over 4}c_3(\omega_\mu \omega^\mu)^2-g_\omega\bar\psi_i \gamma^\mu\psi_i
\omega_\mu\nonumber\\ &-&{1\over 4}\vec B^{\mu\nu}.\vec B_{\mu\nu}
+{1\over 2}m_\rho^2 \vec R^\mu\cdot\vec R_\mu
-g_{\rho}\bar\psi_i\gamma^\mu\vec\tau\psi_i\cdot\vec R^\mu\nonumber\\
 &-&{1\over 4}F^{\mu\nu}F_{\mu\nu}-e\bar\psi_i\gamma^\mu
{\left(1-\tau_{3i}\right)\over 2}\psi_iA_\mu
\label{lag}
\end{eqnarray}

The field for the $\sigma$-meson is denoted by $\sigma$, that of the
$\omega$-meson by $\omega_{\mu}$ and of the isovector $\rho$-meson
by $\vec R_{\mu}$.  $A^{\mu}$ denotes the electromagnetic field.
$\psi_i$ are the Dirac spinors for the nucleons, whose third
component of isospin is denoted by $\tau_{3i}$. Here $g_{s}$,
$g_{\omega}$, $g_{\rho}$ and $e^2/4\pi=1/137$ are
the coupling constants for $\sigma$, $\omega$, $\rho$ mesons and
photon respectively. M is the mass of the nucleon  and $m_{\sigma}$,
$m_{\omega}$ and $m_{\rho}$ are the masses of the $\sigma$, $\omega$
and $\rho$-mesons respectively. $\Omega^{\mu\nu}$,
$\vec{B}^{\mu\nu}$ and $F^{\mu\nu}$ are the field tensors for the
$\omega^{\mu}$, $\vec{\rho}^{\mu}$ and the photon fields
respectively. The field equations for mesons and nucleons are
obtained from the Lagrangian of equation (\ref{lag}) and can be
found in Ref \cite{Gam}. These are nonlinear, coupled partial
differential equations, which are solved self-consistently. These
equations are solved by expanding the upper and lower components
of the Dirac spinors $\psi_i$ and the boson
fields wave functions in terms of a  deformed harmonic oscillator
potential basis.

The total binding energy of the system is
\begin{equation}
E_{tot}=E_{part}+E_\sigma+E_\omega+E_\rho+E_C+E_{pair}+E_{cm}
\end{equation}
where $E_{part}$ is the sum of single particle energies of the nucleons,
$E_\sigma$, $E_\omega$, $E_\rho$ are the contributions of meson energies,
and $E_C$ and $E_{pair}$ are the coulomb and pairing energy respectively.
We have used the pairing gap defined in Ref. \cite{moller} to take pairing
in to account. $E_{cm}=-\frac{3}{4}41 A^{-1/3}$ is the non-relativistic
approximation for the center-of-mass correction.

We use the parameter set TM1, NL3 and NL-SH \cite{param} for our
calculations. We note that in TM1 parameter set has the non-negative
value of the quartic self-coupling coefficient $g_3$ for the omega
mesons. In most of the successful parameter sets the quartic
self-coupling term for sigma meson is negative, so that the energy
spectrum is unbounded below. Although in normal cases the solutions
are obtained in local minimum. However, all these parameter sets
give a good account of various properties such as binding energy,
compressibility, asymmetric energy for nuclear matter.

We have calculated the binding energies, charge radius and shift
$\delta r^2_{p,n}$ and $\delta r^4_{p,n}$ for the barium and radium
isotopes with different parameter sets. The binding energies agree
in all the cases with the experimental values with maximum deviation
of 5 to 6 MeV out of a total binding energy of 1000 MeV for barium
isotopes and similarly for radium isotopes the deviation for binding
energies are around 8 MeV \cite{audi}. The charge radii $r_{ch}$
agree quite well with the fitted values \cite{nadj} for the barium
isotopes to within 1\%. The difference between the root mean squre
radius of neutrons and protons grows almost linearly with the
neutron number for both the barium and radium isotopes. This
difference also depends significantly on the theoretical model used.
The lack of unambiguous precise experimental information on the
neutron distribution means that one must extrapolate to the desired
neutron properties. We note that there is essentially no model
independent experimental information on neutron density
distributions. We next use these radii to estimate the nuclear
structure effects in PNC.

The nuclear structure corrections and the weak charge for different
isotopes of barium and radium evaluated for $S=T=0$ for different
parameter sets. Here one can see that the $q_p$ are constant when
the neutron number increases. However $q_n$ varies slowly as one
increases the neutron number. Our RMF calculation gives the nuclear
structure correction for $^{134}$Ba, $Q_W^{nuc}=3.444$ for TM1,
3.453 for NL3 and 3.436 for NL-SH forces. Similarly the nuclear
structure correction for $^{226}$Ra are $Q_W^{nuc}=15.240$ for TM1,
15.288 for NL3 and 15.215 for NL-SH parameters. In Fig I, we have
plotted the nuclear structure corrections versus different isotopes
of barium (left panel), radium (right panel) for the parameter sets
used in our calculations. It is seen that the Nl3 parameter gives a
higher $Q_W^{nuc}$ compared to other parameters for radium isotopes.
\begin{figure}
\begin{tabular}{cc}
\epsfig{file=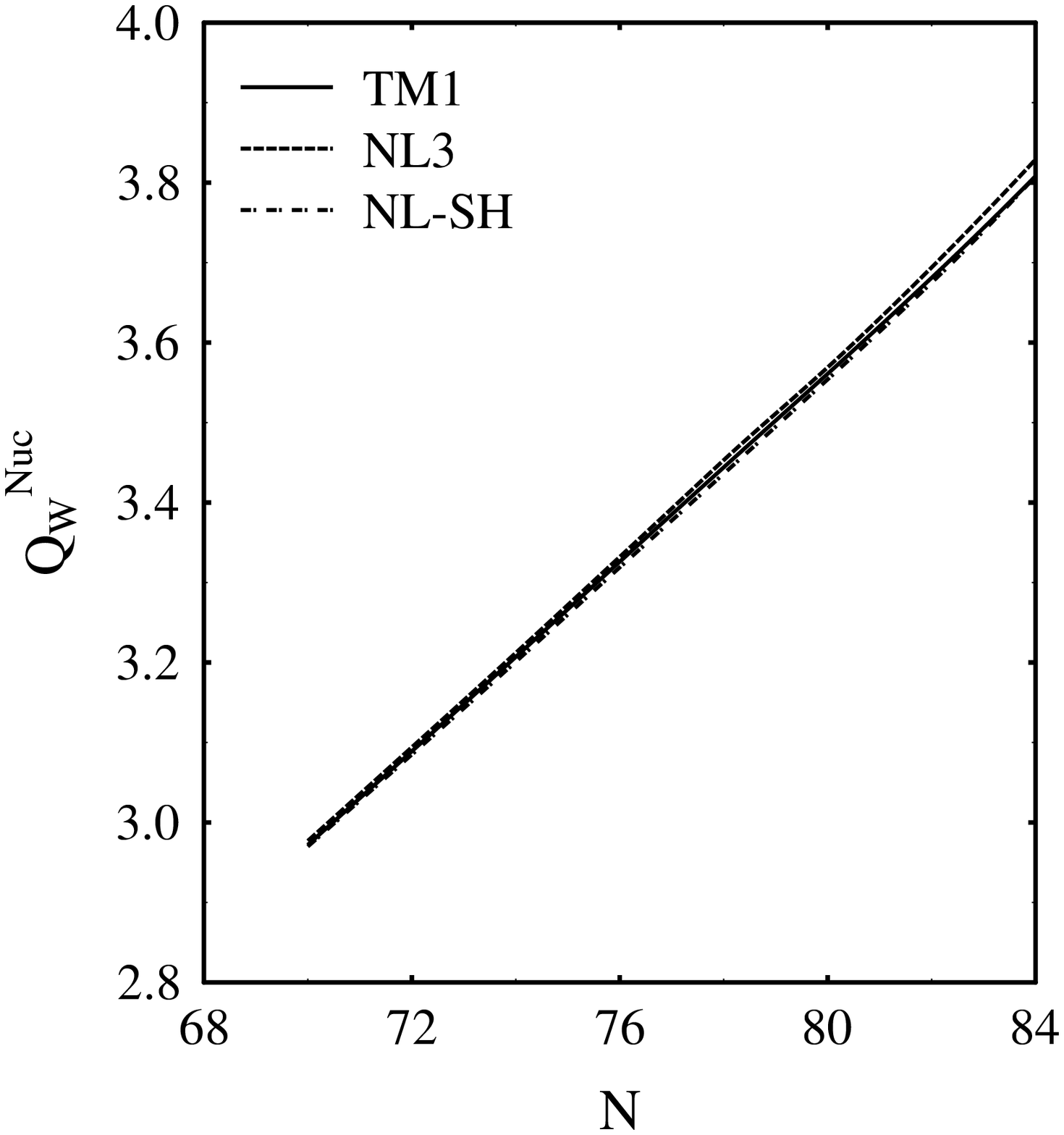,width=4.5cm}&
\epsfig{file=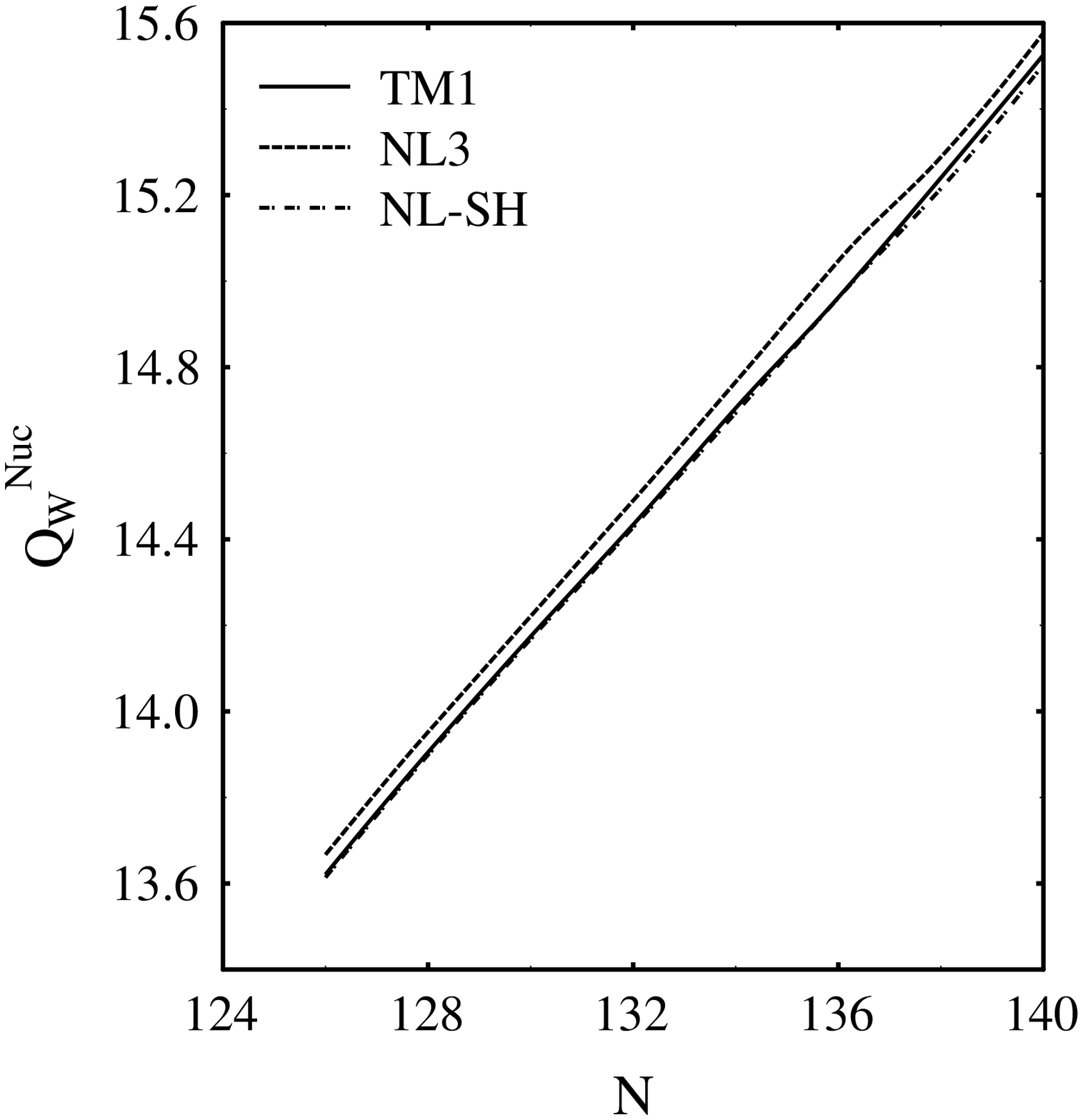,width=4.5cm}\\
\end{tabular}
\caption{N vs. $Q_W^{nuc}$ for barium isotopes(left panel) radium
isotopes (right panel) in different parameter sets}
\end{figure}

We next discuss explicitly the correction to the weak charge arising
from the difference between the neutron and proton distributions.
The small difference between $q_n$ and $q_p$ has the effect of
modifying the effective weak charge as \cite{pollock}
\begin{equation}
Q_W=Q_W^{St. Mod}+\Delta Q_W^{n-p}
\end{equation}
where
\begin{equation}
\Delta Q_W^{n-p}=N(1-q_n/q_p)
\end{equation}
Assuming the difference by a small parameter, $R_n^2/R_p^2=1+\epsilon$,
we have
\begin{equation}
\Delta Q_W^{n-p}\simeq N(Z\alpha)^2(0.221~\epsilon)/q_p
\end{equation}
Our RMF calculation gives $\Delta Q_W^{n-p}=0.252$ in TM1,
$\Delta Q_W^{n-p}=0.260$ in NL3, $\Delta Q_W^{n-p}=0.243$ in NL-SH
parameters for $^{134}$Ba and $\Delta Q_W^{n-p}=1.301$ in TM1,
$\Delta Q_W^{n-p}=1.354$ in NL3 and $\Delta Q_W^{n-p}=1.274$ in NL-SH
parameters for $^{226}$Ra.

In conclusion, we have studied the nuclear weak charges for Ba and
Ra isotopes. Singly charged ions of these atoms have been suggested
for possible measurements of PNC. Our RMF calculation yields $\Delta
Q_W^{n-p}/Q_W$ of 0.35\% for $^{134}$Ba and 1\% for $^{226}$Ra.
These results will have an important bearing on high precision
studies of PNC in a single isotope or a chain of isotopes of Ba$^+$
and Ra$^+$.

\acknowledgements PKP would like to thank the friendly atmosphere at
Indian Institute of Astrophysics, Bangalore, where this work was
partially done.


\begin{thebibliography}{99}
\bibitem{bouch1}M.A. Bouchiat and C. Bouchiat, Phys. Lett. {\bf B 48}, 111
(1974); I.B. Khriplovich, Pis'ma Zh. Eksp. Teor. Fiz {\bf 20}, 686 (1974)
[JETP Lett {\bf 20}, 315 (1974)]; P.G.H. Sandars, {\it Atomic Physics},
edited by G. zu Pulitz (Plenum, New York, 1975) Vol 4, p.71; D.S. Sorede
and E. N. Fortson, Bull, Am. Phys. Soc {\bf 20}, 491 (1975).
\bibitem{edwards}N.H. Edwards, S.J. Phipp, P.E.G. Baird, and S. Nakayama,
Phys. Rev. Lett {\bf 74}, 2654 (1995);D.M. Meekhof, P.K. Majumder, S.K.
Lamoreaux,and E.N. Forston, Phys. Rev. {\bf A 52}, 1895 (1995); M.J.D.
Macpherson, K.P. Zetie, R.B. Warrington, D.N. Stacey, and J.P. Hoare, Phys.
Rev. Lett {\bf 67} 2784 (1991).
\bibitem{wood}C.S. Wood, S.C. Bennett, D. Cho, B.P. Masterson, J.L. Roberts,
C.E. Tanner, and C.E. Wieman, Science, {\bf 275}, 1759 (1997).
\bibitem{dzuba}V.A. Dzuba, V.V. Flambaum, and O.P. Sushkov, Phys. Lett.
{\bf A 141}, 147 (1989).
\bibitem{blundell}S.A. Blundell, W.R. Johnson, and J. Sapirstein, Phys. Rev.
Lett. {\bf 65}, 1411 (1990).
\bibitem{bennett} S.C. Bennett and C.E. Wieman, Phys. Rev. Lett. {\bf 82}, 2484
(1999).
\bibitem{lang}W.J. Marciano, and J.L. Rosner, Phys. Rev. Lett, {\bf 65}, 2963
(1990); P. Langacker, M.-X.Luo, and A.K. Mann, Rev. Mod. Phys. {\bf 64},
87 (1992); J. Erler, and P.Langacker, Euro. J. Phys. {\bf 1}, 90 (1998).
\bibitem{dzuba02}V.A. Dzuba, V.V. Flambaum, and J.S.M Ginges, Phys. Rev.
{\bf D 46}, 076013 (2002); and {\it references therein.}
\bibitem{fortson93}E.N. Fortson, Phys. Rev. lett. {\bf 70}, 2383 (1993).
\bibitem{koerber03}T.W. Koerber, M.H. Schacht, W. Nagourney, and E.N. Fortson,
J. Phys. {\bf B 36}, 637 (2003).
\bibitem{sahoo}B.K. Sahoo, R. Choudhuri, B.P. Das, D. Mukherjee, Paper submitted for
publication; K.P. Geetha, PhD Thesis, Bangalore University, India (2002).
\bibitem{dzuba01}V.A. Dzuba, V.V. Flambaum, and J.S.M Ginges, Phys. Rev.
{\bf A63}, 062101 (2001)
\bibitem{wieman}C.E. Wieman, C. Monroe, and E.A. Cornel, in {\it laser
spectroscopy X} edited by H. Henrikson and P. Vogel (World Scientific,
Singapore, 1990).
\bibitem{fortson}E.N. Fortson, Y. Pang, and L. Wilets, Phys. Rev. Lett.
{\bf 65}, 2857 (1990).
\bibitem{wilets}S.J. Pollock, E.N. Fortson, and L. Wilets, Phys. Rev.
{\bf C 46}, 2587 (1992).
\bibitem{chen}B.Q. Chen and P. Vogel, Phys. Rev. {\bf C 48} 1392 (1993).
\bibitem{panda}P.K. Panda and B.P. Das, Phys. Rev. {\bf C62} 065501 (2000);
D. Vretenar, G.A. Lalazissis, and P. Ring, Phys. Rev. {\bf C62} 045502 (2000).
\bibitem{ginges}J.S.M. Ginges and V.V. Flambaum, Phys. Rep. {\bf 637},
63 (2004).
\bibitem{Teller}H.P. D\"urr, and E. Teller, Phys. Rev. {\bf 101}, 494 (1956).
\bibitem{Johnson}M.H. Johnson, and E. Teller, Phys. Rev. {\bf 98}, 783 (1955).
\bibitem{Duerr}H.P. D\"urr, Phys. Rev. {\bf 103}, 469 (1956).
\bibitem{Walecka}J.D. Walecka, Ann. Phys. (N.Y.) {\bf 83}, 491 (1974).
\bibitem{Set}B.D. Serot, and J.D. Walecka, Adv. Nucl. Phys. {\bf 16}, 1 (1986); Int. J.
Mod. Phys. {\bf E6} 515 (1997).
\bibitem{Gam}Y.K. Gambhir, P. Ring, and A. Thimet, Ann. Phys. (N.Y.) {\bf 198}
132 (1990).
\bibitem{moller}P. Moller {\it et al}, At. Data Nucl. Data Tables {\bf 39}, 225
(1988).
\bibitem{param}Y. Sugahara and H. Toki, Nucl. Phys. {\bf A 579} (1994) 557;
G. A. Lalazissis, J. K\"onig and P. Ring, Phys. Rev. {\bf C 55}, 540
(1997); M.M. Sharma, M.A. Nagrajan and P. Ring, Phys. Lett {\bf B
312} 377 (1993)
\bibitem{audi}G. Audi and A.H. Wapstra, Nucl. Phys. {\bf A 595} (1995) 409.
\bibitem{nadj}E.G. Nadjakov, K.P. Marinova and Yu.P. Gangrsky, At. Data Nucl.
Data Tables {\bf 56}, 133 (1994).
\bibitem{pollock}S.J. Pollock, M.C. Welliver, Phys. Lett {\bf B 464}, 177 (1999).
\end{thebibliography}
\end{document}